# Size dependent nature of the magnetic-field driven superconductor-to-insulator quantum-phase transitions

Xiaofu Zhang,[1]* Adriana E. Lita,[2] Huanlong Liu,[1] Varun B. Verma,[2] Qiang Zhou,[3] Sae Woo Nam,[2] and Andreas Schilling[1]*

[1]*Department of Physics, University of Zürich, Winterthurerstrasse 190, 8057 Zürich, Switzerland.*

[2]*National Institute of Standards and Technology, 325 Broadway, Boulder CO 80305, USA.*

[3]*Institute of Fundamental and Frontier Sciences, University of Electronic Science and Technology of China, Chengdu, Sichuan, China.*

* Corresponding authors, zhang@physik.uzh.ch, schilling@physik.uzh.ch

**The nature of the magnetic-field driven superconductor-to-insulator quantum-phase transition in two-dimensional systems at zero temperature has been under debate since the 1980s, and became even more controversial after the observation of a quantum-Griffiths singularity. Whether it is induced by quantum fluctuations of the superconducting phase and the localization of Cooper pairs, or is directly driven by depairing of these pairs, remains an open question. We herein experimentally demonstrate that in weakly-pinning systems and in the limit of infinitely wide films, a sequential superconductor-to-Bose insulator-to-Fermi insulator quantum-phase transition takes place. By limiting their size smaller than the effective penetration depth, however, the vortex interaction alters, and the superconducting state reenters the Bose-insulating state. As a consequence, one observes a direct superconductor-to-Fermi insulator transition in the zero-temperature limit. In narrow films, the associated critical-exponent products diverge along the corresponding phase boundaries with increasing magnetic field, which is a hallmark of the quantum-Griffiths singularity.**

Superconductivity in strong magnetic fields has been one of the fundamental problems from both the theoretical and practical point of views since its discovery. The superconducting state is a macroscopic quantum state, which is phenomenologically characterized by dissipationless electric currents and the Meissner effect. According to the BCS theory, superconductivity originates microscopically from coherently paired electrons (Cooper pairs)[1]. The superconducting state is therefore characterized by a complex order parameter with an amplitude (related to the energy gap $\Delta$ or the Cooper-pair density $n_s$), and a phase $\phi$. The destruction of the superconducting state in high magnetic fields can occur through the suppression of $n_s$ to zero. However, phase fluctuations of the order parameter can also destroy the zero-resistance state[2]. When the external field exceeds the lower-critical field in type-II superconductors, the magnetic field enters the superconductors via quantized flux lines (vortices). Sufficiently mobile vortices, as a manifestation of phase fluctuations, will generate dissipation and can drive superconductors into normal conductors.

The magnetic-field driven superconductor-to-insulator transition (SIT) in thin films at zero temperature is a well-documented quantum-phase transition[3-5] and has been observed in a myriad of experiments[6-16]. A long-standing controversy is the question whether the SIT is due to the loss of long-range coherence of $\phi$ (bosonic scenario), or due to the breakdown of Cooper pairs that suppresses $n_s$ to zero (fermionic scenario)[3-5]. The bosonic scenario describes the SIT as a result of quantum-phase fluctuations, in which the superconducting and the insulating states with different symmetry are separated by a single quantum-critical point[17,18]. On the superconducting side of the SIT, the Cooper pairs are mobile and the vortices are localized into a vortex lattice or glass, while on the insulating side, the vortices are mobile but the Cooper pairs are localized into isolated superconducting islands, forming a Bose-insulating state[19,20]. In the fermionic scenario, the SIT is driven by the breaking of Cooper pairs and the localization of electrons in high enough fields, forming a Fermi insulator[21-23]. Despite the majority of experiments demonstrate a quantum-fluctuation induced superconductor-to-Bose insulator quantum-phase transition[7-9,14,24,25], a direct superconductor-to-Fermi insulator quantum-phase transition that is induced by the breakdown of Cooper pairs has also been reported[10,11,26].

Moreover, recent observations in crystalline quasi-two-dimensional (2D) and one-dimensional (1D) superconducting systems revealed an unprecedented quantum-Griffiths singularity[27-33], which is experimentally characterized by divergent critical-exponent products of the dynamical critical exponent and the correlation length exponent, $zv$. This novel phenomenon is attributed to quenched-disorder effects at the transitions, despite the origin of such quenched disorder in clean systems is somewhat obscure[27-32,34,35]. The quantum-Griffiths singularity offers a new perspective on the nature of the superconductor-to-insulator or metal-to-insulator transitions, but also evokes a fundamental problem, namely the apparent lack of universality in experiments probing superconductor-to-insulator quantum-phase transitions[3-16,21-33].

Due to the weak pinning in amorphous superconductors, the manifestation of intrinsic vortex interactions can be directly probed, and they therefore represent an optimal platform to study vortex-related phase transitions . Depending on the bridge width $w$, we observe all of the above mentioned magnetic-field induced phenomena in thin amorphous superconducting films, such as superconductor-to-Bose insulator quantum-phase transitions, direct superconductor-to-Fermi insulator phase transitions, and a quantum-Griffiths singularity. Our results can therefore in part explain the seeming absence of universality of these transitions.

## Results and discussion

**Preparation and characterization of the WSi bridges.** We fabricated a series of superconducting microbridges on a single $10 \times 10$ mm$^2$ amorphous WSi thin film with thickness $d = 4$ nm[36,37], spanning the range of bridge widths $w$ over almost three decades from 2 μm to 1000 μm. The amorphous nature of our films is illustrated in Fig. 1a. Standard four-electrode transport measurements taken in zero magnetic field on the as-grown films and on the fabricated long bridges are shown in Fig. 1b. The respective transition temperatures $T_c(0)$ are all $\approx 3.45$ K, demonstrating the homogeneity and quality of the bridges (see Methods section)

**Phase diagram for infinitely wide films.** Figure 1c shows the sheet resistance $R_s(T)$ of the widest 1000-μm-wide bridge at various magnetic fields near the SIT, showing a well-defined quantum-critical field at $B_c^1 = 5.42$ T that separates the zero-temperature superconducting state with $\partial R_s/\partial T > 0$ from the insulating state with $\partial R_s/\partial T < 0$. Above $B_c^1$, superconducting fluctuations still exist on the insulating side up to $B_c^2 \approx 6.4$ T at $T_c(0)$, which supports the scenario of a localization of Cooper pairs induced SIT, i.e., a superconductor-to-Bose insulator quantum-phase transition[14]. The $R_s(T)$ on the insulating side between $B_c^1$ and $B_c^2$ are separated by minima $T_{min}(B)$ into two distinct regimes: a regime with superconducting fluctuations and $\partial R_s/\partial T > 0$ for $T > T_{min}(B)$, and a zero-temperature insulating regime with $\partial R_s/\partial T < 0$ (see Fig. 2a). With increasing temperature, the localized Cooper pairs in this insulating regime finally depair at $T_{min}(B)$, leading to a state with finite-temperature superconducting fluctuations. With increasing magnetic field, the $T_{min}(B)$ plateaus monotonically shift from the zero-temperature quantum-critical point at $B_c^1$ to $T_c(0)$ at $B_c^2$, the highest temperature where Cooper pairs can persist in the 2D superconducting system on the Bose insulating side. Similar results were also observed for the 500-μm-wide bridge (Supplementary Fig. 1b), essentially representing the universal properties of infinite 2D superconducting systems.

**Narrowing down the systems size.** As soon as the bridge width $w$ is further reduced, we fail to identify a quantum-critical regime down to zero temperature. Already for the 200- μm wide bridge (Supplementary Figs. 1c and 2b) the $R_s(T)$ shows a downturn with decreasing temperature near $B_c^1$, instead of a temperature independent value in the zero-temperature limit as we observed it in the wider bridges. By carefully checking all these downturns in $R_s(T)$ shown in Fig. 1 and in the Supplementary Figs. 1 and 2, we find that superconductivity recovers above $B_c^1$ in the narrow bridges, i.e., the initially Bose-insulating state turns into the superconducting state again. When cooling the these bridges down from the normal state (see Supplementary Fig. 3 for the $R_s(T)$ of the 10- and 20-μm wide bridges in 6 T), superconducting fluctuations appear around 6 K with $\partial R_s/\partial T > 0$ at first; then the $R_s(T)$ curves reach a minimum, similar to those in the wide bridges, corresponding to a finite-temperature superconductor-to-Bose insulator phase transition. The Bose insulating state with $\partial R_s/\partial T < 0$, however, does not persist down to the zero-temperature limit, but is terminated by re-entering into the superconducting state with $\partial R_s/\partial T > 0$.

The figures 1d-1f show the $R_s(T)$ of the 50-µm, 10-µm and the 2-µm-wide bridges in the same field range as in Fig. 1c for the 1000-µm-wide bridge, demonstrating that the SIT is completely different from that in the infinite 2D systems. The complete evolution of the SIT as a function of bridge width is shown in the Supplementary Fig. 1 and in Online Video 1. Distinct from the SIT in Fig. 1c, the $R_s(T)$ in the zero-temperature limit drop dramatically with decreasing temperature. The corresponding phase boundaries between the superconducting and the insulating states (Figs. 2b-2d) are well separated by a finite-temperature maximum $T_{max}$ on the $R_s(T)$ curves, and the $dR_s(T)/dT$ changes its sign at a characteristic critical field $B_c^*(T_{max})$ (Supplementary Fig. 3). With increasing $B$, the corresponding $T_{max}$ plateaus shift monotonically down to zero-temperature.

In order to better visualize the recovery of superconductivity in the Bose-insulating state, figures 3 compare representative $R_s(T)$ curves for all the bridges in magnetic fields $\gtrsim B_c^1$ up to 8 T. Despite a $T_{min}$ for fields between $B_c^1$ and $B_c^2$ (Fig. 3a) is observed for all bridges in $B$ = 6 T (i.e., a superconductor-to-Bose insulator transition), the $R_s(T)$ curves show a dramatic decrease in the narrow bridges in the low-temperature limit, suggesting that the freely mobile vortices in the originally Bose-insulating state are re-pinned. By further increasing the magnetic fields above $B_c^2$, the superconductor-to-Bose insulator transition at $T_{min}$ is entirely absent for all bridges. The recovery of superconductivity in the narrow bridges, however, still persists, indicating a direct superconductor-to-Fermi insulator quantum-phase transition in the zero-temperature limit (Figs. 3b and 3c). At sufficiently high fields ($B$ = 8 T), where electrons cannot condense into Cooper pairs even at $T = 0$, the $R_s(T)$ curves do not exhibit any size effects (Fig. 3d).

We have also carefully measured the magnetoresistance $R_s(B)$ at fixed temperatures up to 10 K for all the bridges. In the Supplementary Figs. 4 and 5, we show representative $R_s(B)$ data from 0.35 K to 3.5 K for the 1000-µm- and the 2-µm-wide bridges, respectively, where the phase boundaries are revealed by a series of crossing points, corresponding to the $T_{min}$ and $T_{max}$ plateaus on the $R_s(T)$ curves. For the 1000-µm-wide bridge, the crossing points of these $R_s(B)$ curves gradually shift from $B_c^1 = 5.42$ T at low temperature to $B_c^2 = 6.4$ T at $T_c(0)$, thereby defining the boundary between the superconducting and the Bose-insulating state (see phase diagram in Fig. 2a). For the 2-µm-wide bridge, by contrast, the corresponding crossing points of $R_s(B)$ curves move strongly upward for data below $\approx 1.4$ K, indicating the recovery of superconductivity and an associated bending of the superconductor-to-Bose insulator phase-transition boundary at low temperature (see phase diagrams in Figs. 2b-2d). The evolution of all these phase boundaries with bridge width $w$ are summarized in the Supplementary Figs. 10 and in Online Video 2.

**Scaling analysis along the phase boundaries.** To further investigate the nature of the SITs, we use a scaling analysis on the superconductor-to-insulator phase boundaries. Near the quantum-critical point, physical quantities for an equilibrium system can be classified into distinct universality classes determined only by the general properties of the system, such as space dimensionality, range and dynamics of the interactions, and symmetry, independent of the microscopic details. The experimental characteristics of the quantum-critical state is the scaling behavior of physical quantities within the critical regime, showing a power-law dependence on the rescaled spatial (correlation length $\xi \propto |B - B_c|^{-\nu}$) and temporal (correlation time $\tau \propto \xi^z \propto |B - B_c|^{-z\nu}$) coordinates[17,18].

We first performed the scaling analysis on the 1000-μm-wide bridge, representing the infinitely large case, where the SIT boundary from $B_c^1$ to $B_c^2$ is constituted by a series of $T_{min}$ plateaus (Fig. 4). Near $B_c^1$, on both sides of the transition, the sheet resistance can be rescaled as $R_s(B,T) = R_c F(|B - B_c^1|T^{-1/z\nu})$, where $F(x)$ is a universal scaling function with $F(0) = 1$. By rescaling the field in Fig. 4a as $|B - B_c^1|t$, where $t = T^{-1/z\nu}$, all the $R_s(B,T)$ curves within the quantum-critical regime collapse onto a single curve, as it is shown in Fig. 4b. From the rescaling factor $t$, the critical exponents $z\nu$ can be obtained by a power-law fit of $t(T)$, yielding $z\nu = 1.33 \sim 4/3$ with high precision on both sides of the transition. Similarly, the superconductor-to-Bose insulator quantum-phase transition can also be observed in the 500-μm wide bridge, with the same critical-field value. The resulting $z\nu = 4/3$ has been observed in many 2D superconducting systems, and confirms the universal behaviour of the superconductor-to-Bose insulator quantum-phase transitions in large 2D superconducting systems[7-9,13,14]. Performing a similar scaling analysis for the pair-breaking critical field $B_c^2$ at temperatures between 2.9 K to 3.4 K (Fig. 4c), we obtain the best data collapse with $z\nu = 0.67 \sim 2/3$ (Fig. 4d). These resulting critical-exponent products and the critical behaviour, in general, are consistent with our previous results for infinitely large 2D bridges[14].

As for the narrow bridges, the $B_c^*$ phase boundary corresponds to the $T_{max}(B_c^*)$ plateaus (Figs. 2b-2d and Supplementary Fig. 10). In Fig. 5, we show the corresponding scaling analyses for the 2-μm-wide bridge. The first $T_{max}(B_c^*)$ plateau appears at $B_c^* = 6.1$ T, as it is shown in Fig. 5a, at temperatures between 1.18 K and 1.34 K. By utilizing the scaling analysis, the $R_s(B)$ data within the critical regime are perfectly rescaled onto a single curve, as it is shown in Fig. 5b, resulting in a product of the critical exponents $z\nu = 0.74$. The largest critical point that could be investigated with our equipment is $B_c^* = 7.1$ T (shown in Fig. 5c, at temperatures from 0.34 K to 0.45 K). The best data collapse yields a relatively large critical-exponent product of 2.34 (Fig. 5d). To reveal the effects of the bridge dimensions on the critical behaviour, we performed corresponding scaling analyses every 0.1 T for all the bridges (details about the scaling analysis procedure at the $B_c^*(T_{max})$ boundary are shown in the Supplementary Fig. 6), and the resulting field dependences of $z\nu$ for these bridges are summarized in Fig. 6. Different from the superconductor-to-Bose insulator phase transitions in large 2D systems, in which $z\nu$ lies between 2/3 and 4/3[14], the product $z\nu$ at the $B_c^*(T_{max})$ boundary grows dramatically with increasing magnetic field, and it is expected to diverge at the zero-temperature superconductor-to-Fermi insulator quantum-critical point at $B_c^*(0) = B_c^F$, thereby revealing the signature of a quantum-Griffiths singularity[27-33].

Inspired by the activated scaling law for the quantum-Griffiths singularity[27], we fitted the extracted $z\nu$ for all the bridges according to $z\nu = C \cdot (B_c^F - B_c^*)^{-\nu\Psi}$ (solid lines in Fig. 6; the fitting results are summarized in Table 1 in the Supplementary Information). The resulting critical-exponent products $\nu\Psi$ are found to strongly depend on the bridge width, which can be attributed to the different quenched disorder levels among different bridges as we will outline below. The most remarkable result of this fitting procedure is that the resulting $B_c^F = B_c^*(0)$ values for all bridges are around $7.25 \pm 0.01$ T. This size-independent field $B_c^F$ represents the highest magnetic field in which Cooper pairs can still exist in the films in the zero-temperature limit. It therefore corresponds to the zero-temperature quantum-critical field for the direct superconductor-to-Fermi insulator quantum-phase transitions for narrow enough bridges, irrespective of the geometries of the superconducting structures.

**Decisive role of the system size.** The physical origin of the seemingly contradictory nature of the phase transitions in the different bridges can be well understood in the context of vortex physics. In superconducting thin films, vortices interact via their stray fields in the surrounding space, different from their bulk peers[38]. The stray field in a 2D system is mainly mediated by the effective penetration depth $\Lambda = 2\lambda_L^2/d$, with $\lambda_L$ the London penetration depth, which is $\Lambda(0) \approx 350$ μm in our case[36,37]. For an infinitely large 2D superconducting thin film in a strong magnetic field, vortices localize due to the long-range logarithmic repulsive interactions $V_{int} \propto \ln(\Lambda/r)$, where $r$ is the distance between two vortices. As a result, field-induced vortices freeze into a regular vortex lattice in the zero-temperature limit, underlying the global coherence, and the zero-resistivity state survives at first in a strong magnetic field. When the magnetic field is increased further, the vortices eventually condense (i.e., delocalize) at a critical field $B_c^1$, in analogy to the condensation of Cooper pairs and the localization of vortices in the superconducting state[17,18]. Although global superconducting coherence is then destroyed, isolated superconducting islands remain, with mobile vortices that are induced by quantum phase fluctuations[17-20]. This transition can be regarded as a quantum analog of the vortex-unbinding induced Kosterlitz-Thouless transition. In short, the quantum nature of the superconductor-to-Bose insulator quantum-phase transition in sufficiently wide structures corresponds to the delocalization of vortices and the localization of Cooper pairs, accompanied by the loss of the global coherence[17,18].

The interaction between vortices can, however, change from long-range to an exponentially weak and short-range interaction as soon as the characteristic length scale of the superconducting system is smaller than $\Lambda(0)$[39,40]. This may prohibit the formation of a long-range ordered vortex lattice. As we recently demonstrated in similar superconducting microbridges[37], an extra pinning effect for bridge widths smaller than $\Lambda(0)$ prohibits the vortices from moving, leading to a strong suppression of the resistivity even near the normal-to-superconductor transition at the critical temperature $T_c(B)$ (see also Supplementary Fig. 7). Upon approaching the zero-temperature limit, where thermal fluctuation effects on the vortices are negligible, the initially mobile vortices in the Bose-insulating state freeze again due to this size effect in the narrow bridges, which leads to a dramatic decrease in resistivity and the recovery of superconductivity. Within this line of arguments, vortices do not constitute a long-range ordered vortex lattice in bridges with dimensions smaller than $\Lambda(0)$. Instead, they are prone to form locally ordered regions, which are connected by disordered and randomly located vortices with quenched disorder, leading to a vortex-glass-like state near $B_c^F$, as schematically illustrated in ref. 33. This state breaks the transitional and rotational symmetry of the vortex lattice, and dramatically alters the nature of the transition, manifesting itself in a quantum-Griffiths singularity[34]. A direct superconductor-to-Fermi insulator quantum-phase transition due to the direct Cooper-pair breaking eventually occurs at the critical field $B_c^F$ in the zero-temperature limit.

Our experimental findings are summarized in the schematic $B$-$T$ phase diagrams in Fig. 2 (see Supplementary Figs. 8 and 9 for a summary of the corresponding $R_s(T)$ data for different magnetic fields; the Supplementary Fig. 10 and Online Video 2 summarize the phase diagrams for all bridges). The Fig. 2a represents the sequential superconductor-Bose insulator-Fermi insulator quantum-phase transitions for infinite films $w \gg \Lambda(0)$, where $B_c^1$ is the critical field for the localization of Cooper pairs and the $B_c^2$ is the highest field in which Cooper pairs can persist in the films at $T_c(0)$. The ultimate pair-breaking field of localized pairs in the Bose-insulating state is represented by the dashed line. As soon as the bridge width is narrower than the Pearl length $\Lambda(0)$, but sill much larger than the coherence length $\xi(0) \approx 8$ nm[37], superconductivity is recovered from the Bose-insulating state, forming a direct superconductor-to-Fermi insulator quantum-phase transitions in the zero-temperature limit. At finite temperatures $T > 0$, the Bose-insulating phase shrinks with decreasing bridge width, terminating at $B_c^F = 7.25$ T, but it exists up to $T_c(0)$ at $B_c^2$ for all bridges (see Supplementary Fig. 11). By further reducing the bridge width towards the one dimensional limit, however, we expect that the Bose-insulating state will eventually entirely disappear (Fig. 2e), so that the phase diagram is divided into only two distinct regions (with $\partial R_s/\partial T > 0$ and $\partial R_s/\partial T < 0$, respectively), as it is observed for a quantum-Griffiths singularity[27-33]. A direct superconductor-to-Fermi insulator phase transitions occurs at all temperatures up to $T_c(0)$ and beyond as long as superconducting fluctuations are still present.

## Conclusion

We have experimentally demonstrated that in infinite 2D superconducting systems with intrinsically weak vortex pinning, a superconductor-to-Bose insulator quantum-phase transition occurs, which can be attributed to the condensation of vortices. The recovery of superconductivity and the appearance of a quantum-Griffiths singularity in the zero-temperature limit in narrow bridges are most probably due to size-effect induced disorder and a resulting breaking of transitional and rotational symmetry of the vortex lattice. The quantum-Griffiths singularity should therefore be universally expected in sub-2D superconducting systems with characteristic dimensions smaller than $\Lambda(0)$. Our experimental findings may solve part of the controversy concerning the nature of magnetic-field driven superconductor-to-insulator quantum-phase transitions in 2D superconducting systems.

# Online content/Additional information

Supplementary Information: Supplementary Figures 1-11, Table 1, Online Videos 1-2

# References

1. Bardeen J., Cooper L. N., Schrieffer J. R., Theory of superconductivity. Phys. Rev. **108**, 1175–1204 (1957).
2. Rasolt M., Tešanović Z., Theoretical aspects of superconductivity in very high magnetic fields. Rev. Mod. Phys. **64**, 709–754 (1992).
3. Goldman, A. M. Superconductor-insulator transitions. Int. J. Mod. Phys. B **24**, 4081–4101 (2010).
4. Gantmakher, V. F. & Dolgopolov, V. T. Superconductor–insulator quantum phase transition. Phys. Usp. **53**, 1–49 (2010).
5. Sacépé B., Feigel'man M., Klapwijk T. M., Quantum breakdown of superconductivity in low-dimensional materials. Nat. Phys. **16**, 734–746 (2020).
6. Haviland, D. B., Liu, Y. & Goldman, A. M. Onset of superconductivity in the two-dimensional limit. Phys. Rev. Lett. **62**, 2180–2183 (1989).
7. Hebard, A. F. & Paalanen, M. A. Magnetic-field-tuned superconductor–insulator transition in two-dimensional films. Phys. Rev. Lett. **65**, 927–930 (1990).
8. Yazdani, A. & Kapitulnik, A. Superconducting–insulating transition in two-dimensional a-MoGe thin films. Phys. Rev. Lett. **74**, 3037–3040 (1995).
9. Mason N., & Kapitulnik N. Dissipation effects on the superconductor-insulator transition in 2D superconductors. Phys. Rev. Lett. **82**, 5341-5344 (1999).
10. Marković, N., Christiansen, C. & Goldman, A. Thickness-magnetic field phase diagram at the superconductor–insulator transition in 2D. Phys. Rev. Lett. **81**, 5217–5220 (1998).
11. Aubin, H. et al. Magnetic-field-induced quantum superconductor–insulator transition in $Nb_{0.15}Si_{0.85}$. Phys. Rev. B **73**, 094521 (2006).
12. Biscaras, J. et al. Multiple quantum criticality in a two-dimensional superconductor. Nat. Mater. **12**, 542–548 (2013).
13. Shi, X., Lin, P. & Sasagawa, T. Two-stage magnetic-field-tuned superconductorinsulator transition in underdoped $La_{2-x}Sr_xCuO_4$. Nat. Phys. **10**, 437–443 (2014).
14. Zhang, X., Schilling, A. Sequential superconductor–Bose insulator–Fermi insulator phase transitions in quasi-two-dimensional $a$-WSi. Phys. Rev. B **97**, 214524 (2018).
15. Lu, X., et al. Superconductors, orbital magnets and correlated states in magic-angle bilayer graphene. Nature **574**, 653-657 (2019).
16. Yu, Y. et al. High-temperature superconductivity in monolayer $Bi_2Sr_2CaCu_2O_{8+\delta}$. Nature **575**, 156–163 (2019).
17. Fisher, M. P. A., Weichman, P. B., Grinstein, G., Fisher, D. S. Phys. Rev. B **40**, 546-570 (1989).
18. Fisher, M. P. A., Quantum phase transitions in disordered two-dimensional superconductors. Phys. Rev. Lett. **65**, 923-926 (1990).
19. Phillips P., & Dalidovich D. The elusive Bose metal. Science **302**, 243-247 (2003).
20. Phillips P., Two-dimensional materials: not just a phase. Nat. Phys. **12**, 206-207 (2016).

21. Valles, J. M., Dynes, R. C. & Garno, J. P. Electron tunneling determination of the order-parameter amplitude at the superconductor-insulator transition in 2D. Phys. Rev. Lett. **69**, 3567–3570 (1992).
22. Hsu, S. Y., Chervenak, J. A., & Valles Jr, J. M. Magnetic field enhanced order parameter amplitude fluctuations in ultrathin films near the superconductor-insulator transition. Phys. Rev. Lett. **75**, 132-135 (1995).
23. Szabó, P. et al. Fermionic scenario for the destruction of superconductivity in ultrathin MoC films evidenced by STM measurements. Phys. Rev. B **93**, 014505 (2016).
24. Allain, A., Han, Z. & Bouchiat, V. Electrical control of the superconducting-to-insulating transition in graphene-metal hybrids. Nature Mater. **11**, 590–594 (2012).
25. Bollinger, A. T. et al. Superconductor–insulator transition in $La_{2-x}Sr_xCuO_4$ at the pair quantum resistance. Nature **472**, 458–460 (2011).
26. Caviglia, A. D. et al. Electric field control of the $LaAlO_3$/$SrTiO_3$ interface ground state. Nature **456**, 624–627 (2008).
27. Xing, Y. et al. Quantum Griffiths singularity of superconductor-metal transition in Ga thin films. Science **350**, 542 (2015).
28. Xing, Y. et al. Ising superconductivity and quantum phase transition in macrosize monolayer $NbSe_2$. Nano Lett. **17**, 6802 (2017).
29. Shen, S. et al. Observation of quantum Griffiths singularity and ferromagnetism at the superconducting LaAlO3/SrTiO3(110) interface. Phys. Rev. B **94**, 144517 (2016).
30. Saito, Y., Nojima, T. & Iwasa, Y. Quantum phase transitions in highly crystalline two-dimensional superconductors. Nat. Commun. **9**, 778 (2018).
31. Liu, Y., et al. Anomalous quantum Griffiths singularity in ultrathin crystalline lead films. Nature communications **10**, 1-6 (2019).
32. Zhang, E. et al. Signature of quantum Griffiths singularity state in a layered quasi-one-dimensional superconductor. Nat. Commun. **9**, 4656 (2018).
33. Lewellyn, N. A. et al. Infinite-randomness fixed point of the quantum superconductor-metal transitions in amorphous thin films. Phys. Rev. B **99**, 054515 (2019).
34. Markovic, N. Randomness rules. Science **350**, 509–509 (2015).
35. Saito, Y., Nojima, T. & Iwasa, Y. Highly crystalline 2D superconductors. Nat. Rev. Mater. **2**, 16094 (2016).
36. Zhang, X. et al. Superconducting fluctuations and characteristic time scales in amorphous WSi. Phys. Rev. B **97**, 174502 (2018).
37. Zhang, X. et al. Strong suppression of the resistivity near the superconducting transition in narrow microbridges in external magnetic fields. Phys. Rev. B **101**, 060508 (2020).
38. Pearl, J. Current distribution in superconducting films carrying quantized fluxoids. Appl. Phys. Lett. **5**, 65-66 (1964).
39. Kogan, V. G. Pearl's vortex near the film edge. Phys. Rev. B **49**, 15874-15878 (1994).
40. Kogan, V. G. Interaction of vortices in thin superconducting films and the Berezinskii-Kosterlitz-Thouless transition. Phys. Rev. B **75**, 064514 (2007).

## Methods

The superconducting thin films adopted in our research were prepared by magnetron sputtering deposition. The WSi films were deposited by co-sputtering from W and Si targets in 1.2 mTorr Ar pressure, on an oxidized Si substrate. The sputtering powers for W and Si guns were 100 W and 180 W, respectively. The WSi film was *in situ* capped with a 2 nm sputtered amorphous Si film. The WSi film had a nominal Si content of ~ 25% and showed an amorphous structure as verified by X-ray diffraction and Cross-sectional Transmission Electron Microscopy.

We at first patterned Ti/Au contacts on the as-grown films by lift-off technique. Then the micro-bridges were defined by optical lithography, followed by reactive ion-etching. The bridge widths range from 2 µm to 1000 µm. In order to exclude to any possible formation of pinning centers by high-energy electron irradiation, we here only applied optical lithography to fabricate these bridges instead of electron-beam lithography. The bridge length was 4000 µm and 3000 µm long for the 1000- and the 500-µm-wide bridges, respectively. For all other bridges, the length was 800 µm. Images of similar micro-bridges can be found in ref. 36. The resistivity measurements were done in a Physical Property Measurement System (PPMS, *Quantum Design Inc.*) equipped with a $^3$He option. In order to make the results for the different bridge widths comparable, all measurements were performed with same low current density $j = 1.25$ MA/m$^2$, which is around four orders of magnitude below the depairing critical-current density. The bias currents were therefore ranging from 10 nA (for the 2-µm-wide bridge) to 5 µA (for the 1000 µm-wide bridge).

Weakly-pinning amorphous superconductors, such as WSi, InO$_x$, and MoGe, are the most optimal platforms for investigating the quantum nature of superconductor-to-insulator quantum phase transitions, the details of which depend on the intrinsic interactions among vortices. In the strongly-pinning peers, such as NbN, the pinning centers inside the materials prevent vortices from freely moving, making the interactions between free vortices inaccessible. Although atomically ordered superconducting crystalline films such as exfoliated NbSe$_2$ monolayers would also be good candidates. The generally available thin flakes usually have limited size, however, and the effective penetration depth can be orders of magnitude larger than the flake size, so that the limit $w \gg \Lambda(0)$ is hardly accessible.

**Acknowledgements:** Funding: Q.Z. acknowledges funding from the National Key R&D Program of China (2018YFA0307400) and the National Natural Science Foundation of China under grants 61775025 and 61405030. H.L. acknowledges funding from the Swiss National Foundation (20-175554).

**Author contributions:** X.Z. and A.S. conceived the research. A.E.L., V.B.V., and S.W.N. deposited the films. X.Z. fabricated the devices and performed all the measurements. X.Z., H.L., Q.Z. and A.S. analyzed the data and wrote the paper. Competing interests: The authors declare that they have no competing interests. Data and materials availability: All data needed to evaluate the conclusions in the paper are present in the paper and/or the Supplementary Information. Additional data related to this paper may be requested from the authors.

**Competing interests:** The authors declare no competing interests.

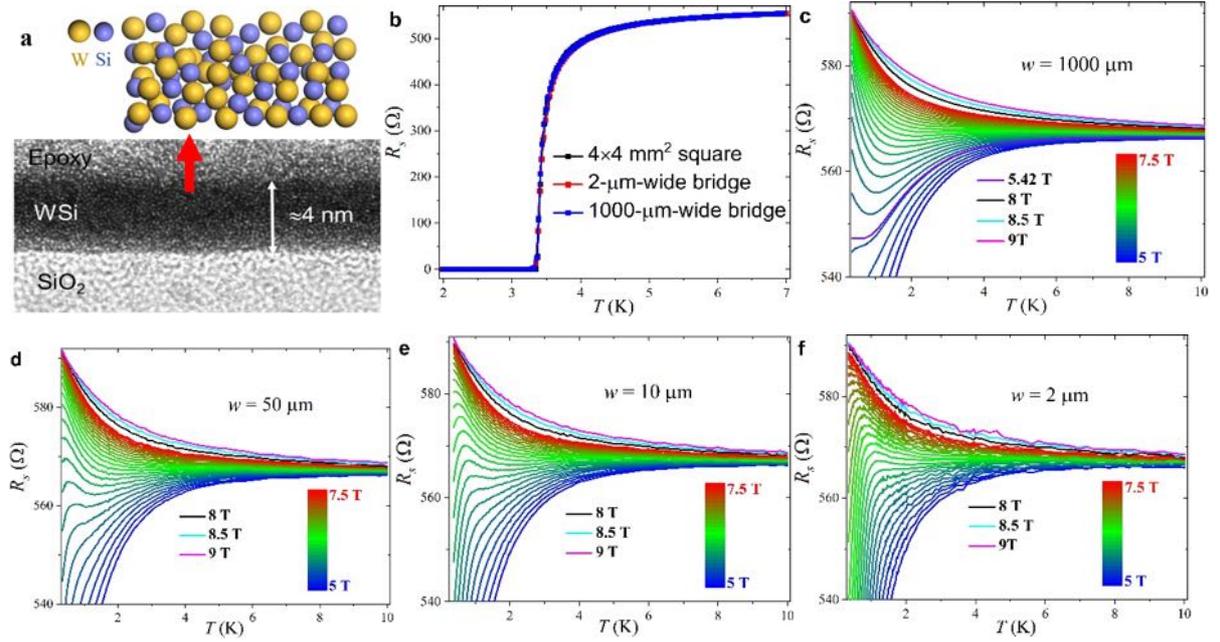

**Fig. 1 | Superconductor-to-insulator quantum-phase transitions**. (**a**) Schematic view of the amorphous WSi films. The amorphous nature and the thickness is examined by cross-sectional transmission-electron microscopy. (**b**) Zero-field normal-to-superconductor transitions of the bare films and the microbridges. (**c**) $R_s(T)$ in magnetic fields from $B$ = 5 to 9 T of the 1000-μm-wide bridge. The violet line shows the separatrix at the quantum-critical field $B_c^1$ for the superconductor-to-Bose insulator quantum-phase transition. (**d-f**) Corresponding $R_s(T)$ data of the 50μm, 10 μm and the 2-μm-wide bridges, respectively.

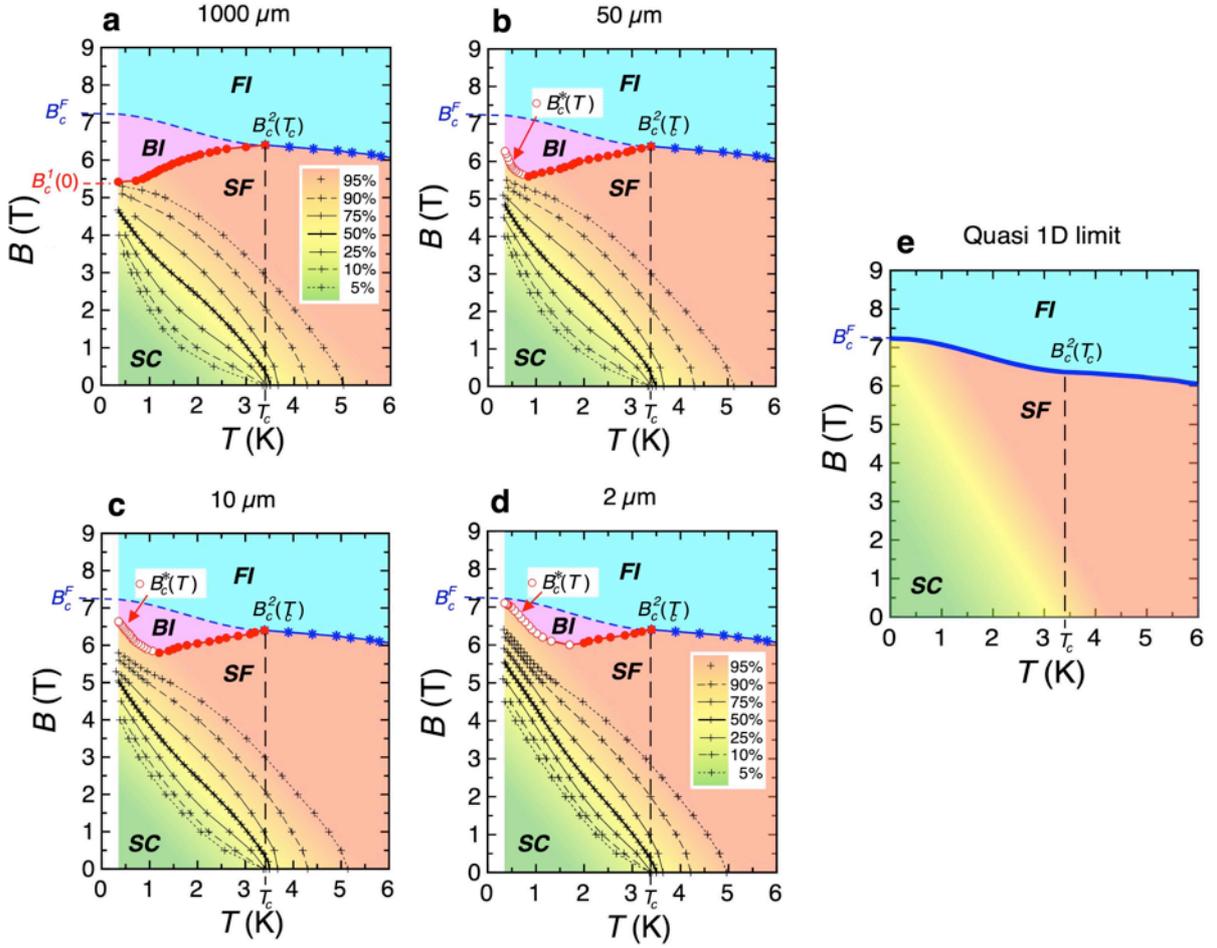

**Fig. 2 | The schematic *B-T* phase diagrams.** (**a**) The sequential superconductor (SC)-Bose insulator (BI) -Fermi insulator (FI) quantum-phase transitions for infinite 2D superconducting systems. (**b** to **d**) Superconductor-to-insulator phase transitions for quasi-2D superconducting systems with dimensions smaller than $\Lambda(0)$. The lines connecting the crosses (+) in (**a** to **d**) represent the measured resistance values for fixed fractions of the normal-state resistance from 5% to 95%, extending to the region dominated by superconducting fluctuations (SF). (**e**) Expected superconductor-to-insulator phase transition for systems approaching the 1D limit.

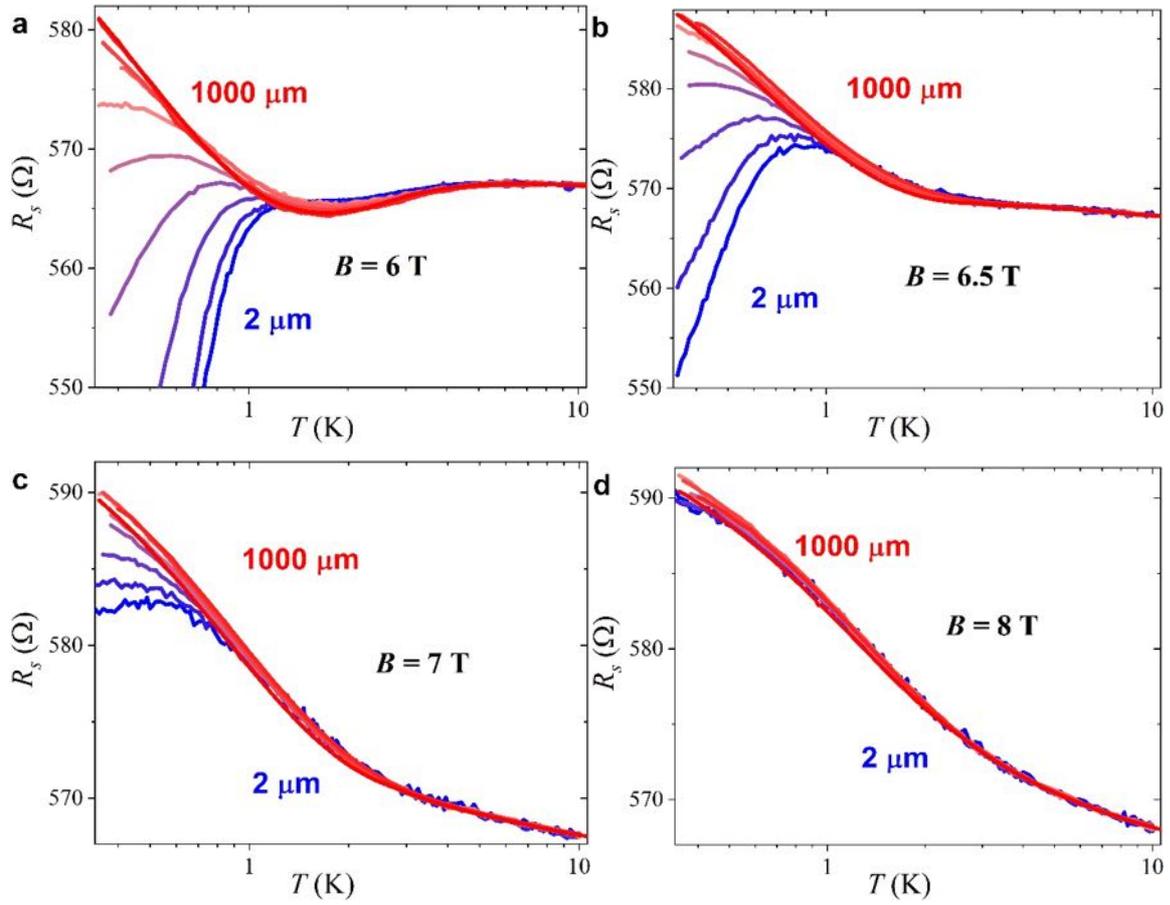

**Fig. 3 | Recovery of superconductivity in magnetic fields between $B_c^1 = 5.42$ T and $B_c^F = 7.25$ T upon decreasing the bridge width** (**a** to **d**) The $R_s(T)$ curves of all the bridges taken in $B = 6$ T, 6.5 T, 7 T and 8T, respectively.

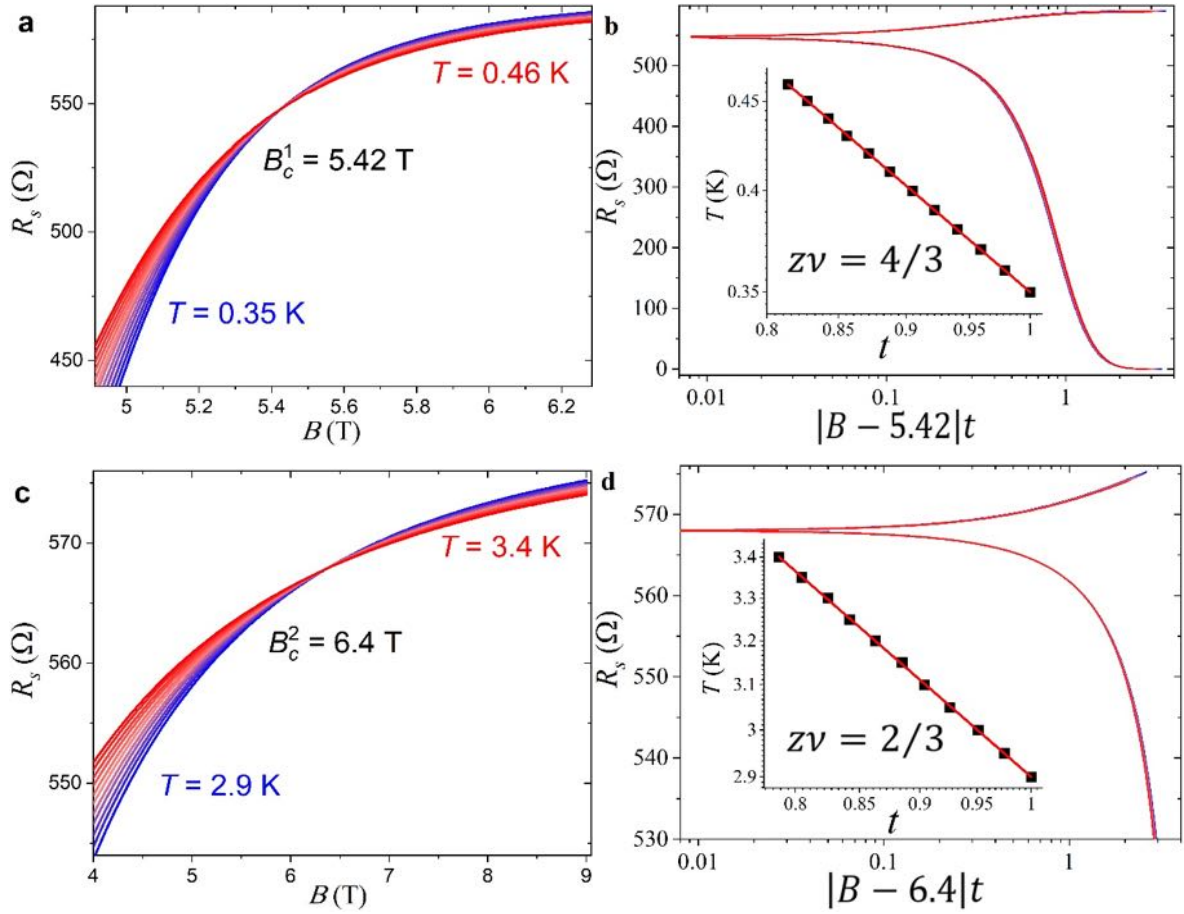

**Fig. 4 | Scaling analysis on the superconductor-to-Bose insulator quantum phase transitions of the 1000-μm-wide bridge**. (**a**) Sheet resistance $R_s$ as a function of magnetic field at temperatures from 0.35 to 0.46 K, showing a distinct crossing point at $B_c^1$. (**b**) Scaling-analysis plot of $R_s$ as a function of $|B - B_c^1|t$. Inset: temperature dependence of the scaling parameter $t$, with $z\nu = 4/3$. (**c**) $R_s(B)$ curves at temperatures from 2.9 to 3.4 K, manifesting the pair-breaking critical field $B_c^2$ near $T_c(0)$. (**d**) Scaling-analysis plot within the critical regime at $B_c^2$. Inset: temperature dependence of the scaling parameter $t$, with $z\nu = 2/3$.

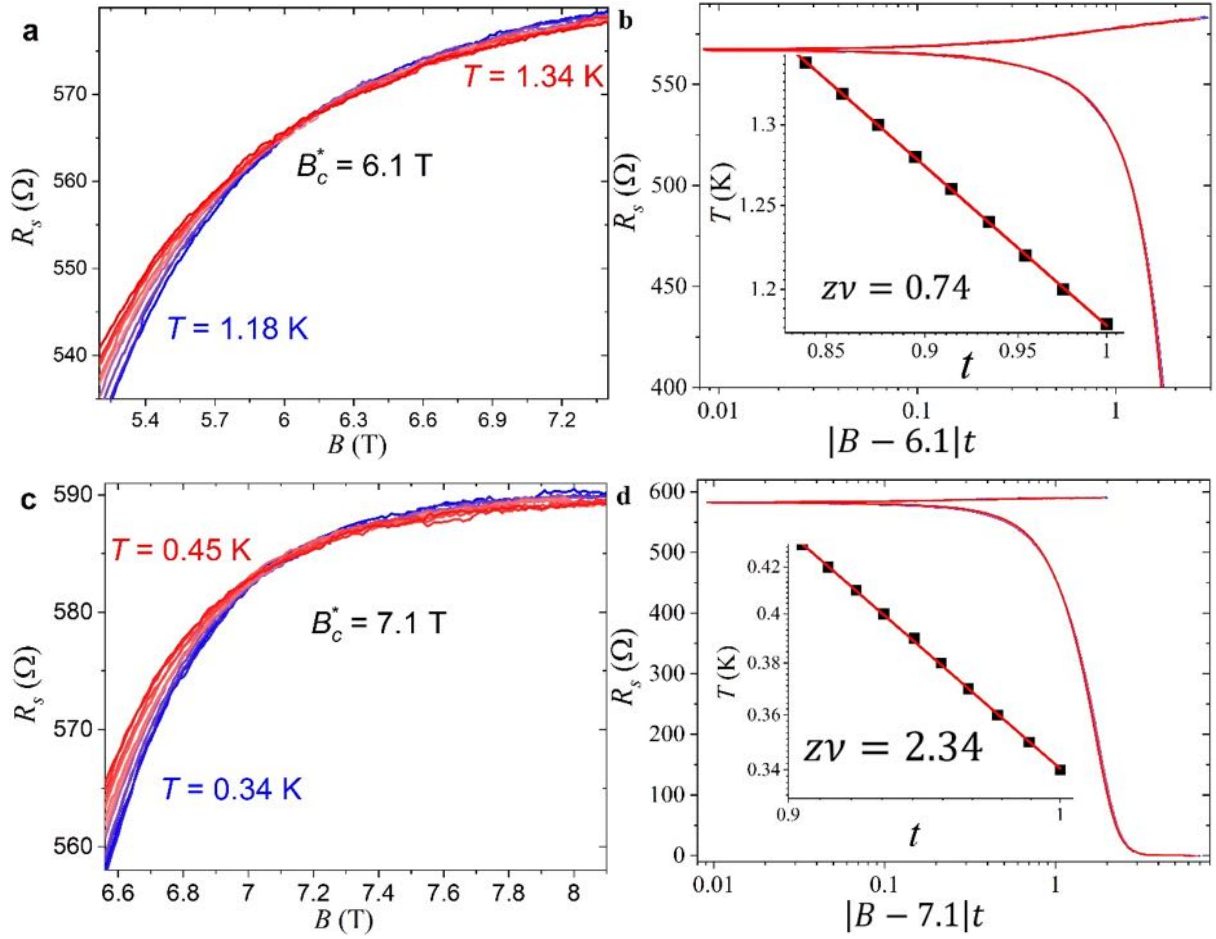

**Fig. 5 | Scaling analysis on the superconductor-to insulator quantum phase transitions of the 2-μm-wide bridge**. (**a**) Sheet resistance $R_s$ as a function of magnetic field at temperatures between 1.18 and 1.34 K, showing a distinct crossing point of the first $T_{max}$ plateau at $B_c^* = 6.1$ T. (**b**) Scaling-analysis plot of $R_s$ as a function of $|B - 6.1|t$. Inset: temperature dependence of the scaling parameter $t$, with $zv = 0.7365$. (**c**) Sheet resistance as a function of field $R_s(B)$ at temperatures between 0.34 and 0.45 K, showing the crossing point of the $T_{max}$ plateau at $B_c^* = 7.1$ T. (**d**) Scaling-analysis plot within the critical regime at 7.1 T. Inset: temperature dependence of the scaling parameter $t$, with $zv = 2.3443$.

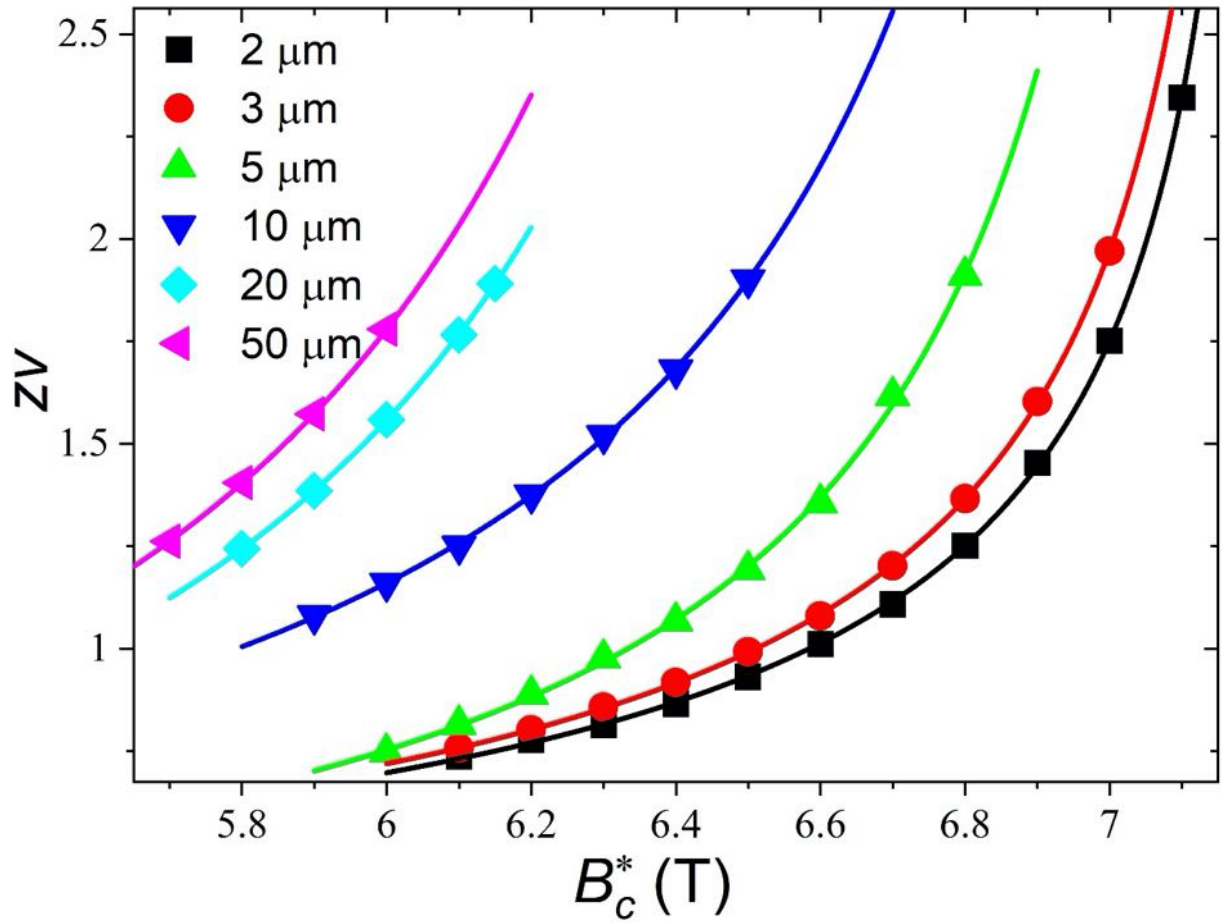

**Fig. 6 | The scaling behaviour of *zv* as a function of critical field $B_c^*$ for narrow bridges.** The solid lines represent fits according to the activated scaling law $zv = C \cdot (B_c^F - B_c^*)^{-v\Psi}$.

# Supplementary Information for "Size dependent nature of the magnetic-field driven superconductor-to-insulator quantum-phase transitions"

(see **https://doi.org/10.1038/s42005-021-00602-7** for the final version with updated figures)

Xiaofu Zhang,[1]† Adriana E. Lita,[2] Huanlong Liu,[1] Varun B. Verma,[2] Qiang Zhou,[3] Sae Woo Nam,[2] and Andreas Schilling[1]†

[1] *Department of Physics, University of Zürich, Winterthurerstrasse 190, 8057 Zürich, Switzerland.*

[2] *National Institute of Standards and Technology, 325 Broadway, Boulder CO 80305, USA.*

[3] *Institute of Fundamental and Frontier Sciences, University of Electronic Science and Technology of China, Chengdu, Sichuan, China.*

† Corresponding authors, zhang@physik.uzh.ch, schilling@physik.uzh.ch

**Supplementary Figures 1-11**

Fig. 1. Evolution of the sheet resistivity around the superconductor-to-insulator quantum phase transitions as a function of the bridge width.

Fig. 2. Recovery of superconductivity in the Bose insulating state.

Fig. 3. Normal-to-superconductor transitions in strong magnetic field.

Fig. 4. Superconductor-to-insulator phase boundary for the 1000-$\mu$m-wide bridge.

Fig. 5. Superconductor-to-insulator phase boundary for the 2-$\mu$m-wide bridge.

Fig. 6. Schematic diagram for the scaling analysis for the superconductor-to-insulator transition at $T_{max}(B_c^*)$.

Fig. 7. The size effect induced suppression of the resistivity near the normal-to-superconductor transition.

Fig. 8. Illustrating diagrams for the magnetic phase diagram of the superconductor-to-Bose insulator transition.

Fig. 9. Illustrating diagrams for the magnetic phase diagram in the narrow bridges.

Fig. 10. Schematic *B-T* phase diagrams for all the bridges.

Fig. 11. Demonstration of the depairing critical field $B_c^2$ at $T_c(0)$ for the 2-, 10-, and 1000-$\mu$m-wide bridges.

**Supplementary Table 1.** Fitting results from the activated scaling law for all bridges.

**Online Video 1** for a clearer visualization of the evolution of $R_s(B,T)$ with the bridge width.

**Online Video 2** showing the evolution of the magnetic phase diagrams as a function of bridge width.

**Supplementary Figures**

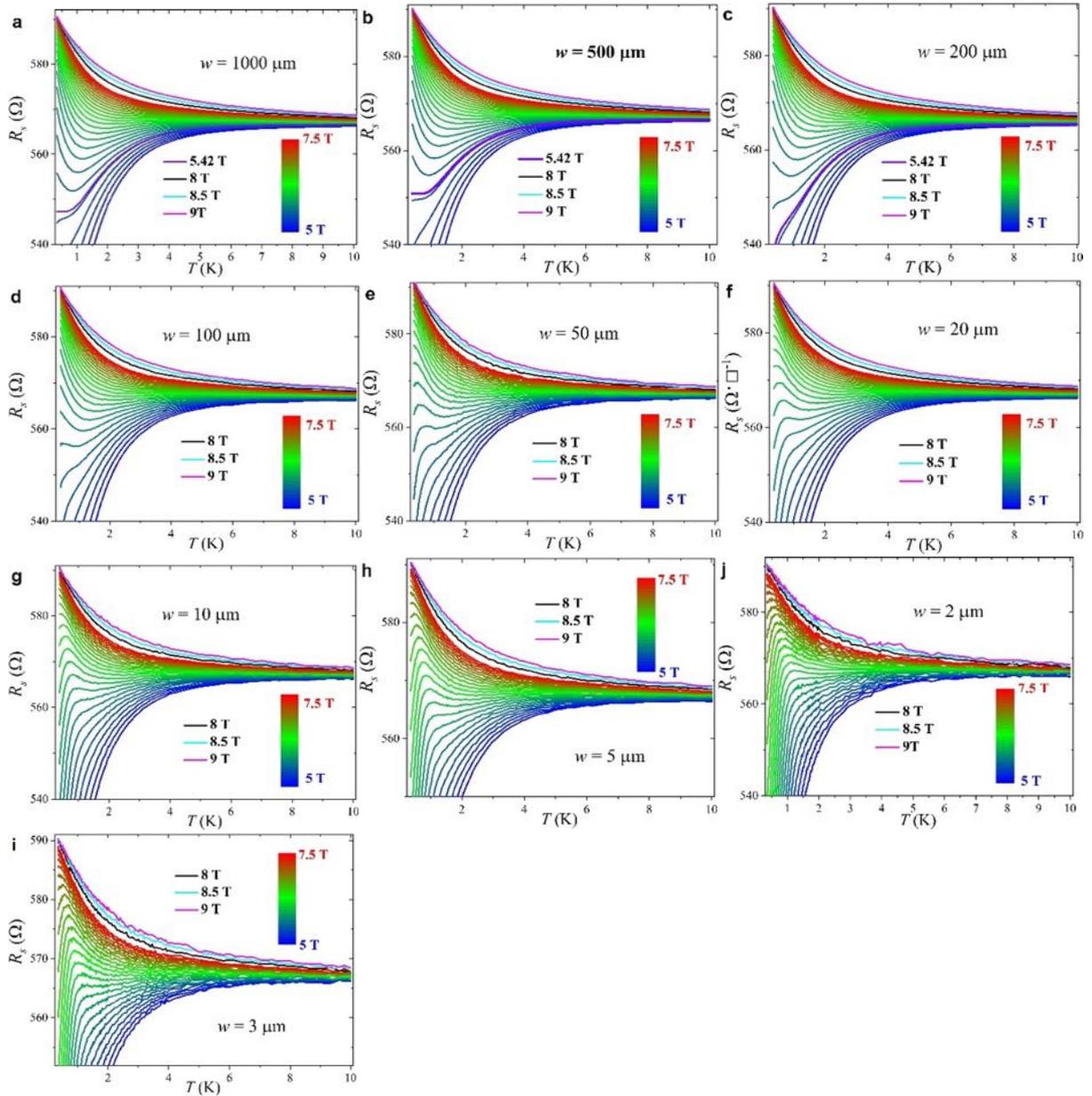

**Fig. 1 | Evolution of the sheet resistivity around the superconductor-to-insulator quantum phase transitions as a function of the bridge width.** (**a** to **c**) Sheet resistance $R_s$ as a function of temperature in magnetic fields ranging from 5 to 9 T for the 1000-, 500-, and 200-$\mu$m-wide bridges, respectively. The violet line corresponds to the Cooper-pair localization quantum critical field $B_c^1 = 5.42$ T. (**d** to **j**) $R_s(T)$ curves in magnetic fields from 5 to 9 T for the 100-, 50-, 20-, 10-, 5-, and 3-$\mu$m-wide bridges, respectively. For a clearer visualization of the evolution of $R_s(B,T)$ on the bridge width, see the Online Video 1 on https://www.physik.uzh.ch/groups/schilling/paper/Resistivity.mov

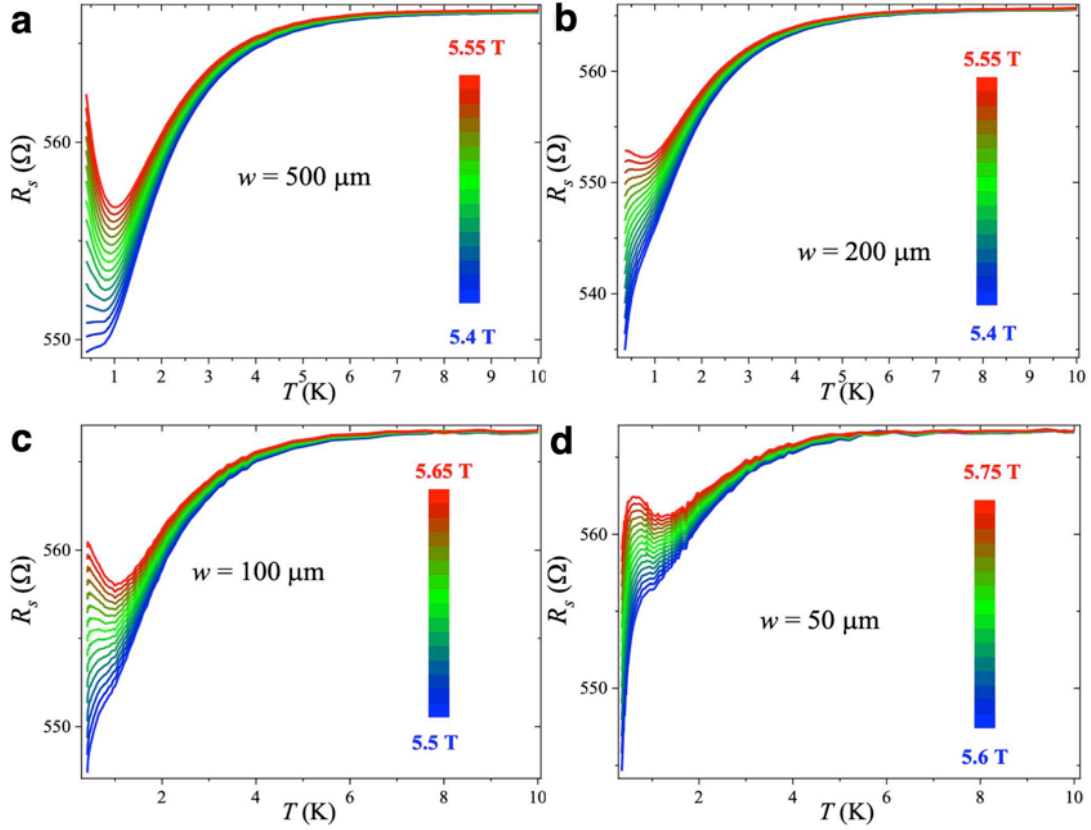

**Fig. 2 | Recovery of superconductivity in the Bose insulating state.** (**a**) The $R_s(T)$ data in magnetic fields from 5.4 to 5.55 T for the 500-µm-wide bridge, showing a clearly insulating behaviour for $B \gtrsim B_c^1 \sim 5.42$ T. (**b**) The $R_s(T)$ curves near $B_c^1$ for the 200-µm-wide bridge. A temperature independent $R_s$ at $B_c^1$ down to the zero-temperature limit is absent. (**c** and **d**) The $R_s(T)$ curves near the SIT for the 100- and 50-µm-wide bridges, respectively. Superconductivity is clearly recovered for $B > B_c^1$ from the previous Bose insulating state.

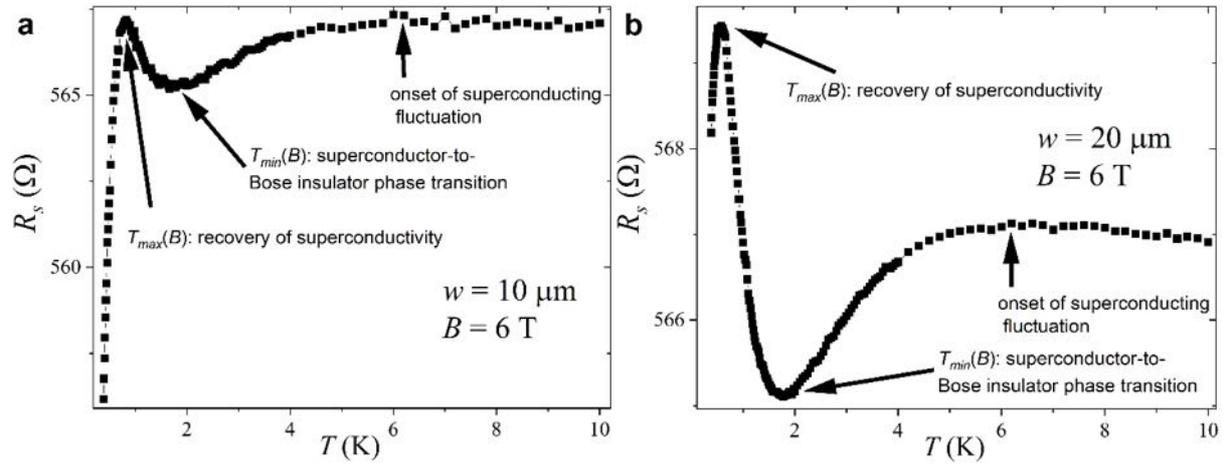

**Fig. 3 | Normal-to-superconductor transitions in strong magnetic field.** (**a** and **b**) The sheet resistance $R_s(T)$ from 0.35 K to 10 K in $B = 6$ T for the 10- and 20-μm-wide bridges, respectively. We have marked all the relevant critical temperatures with arrows.

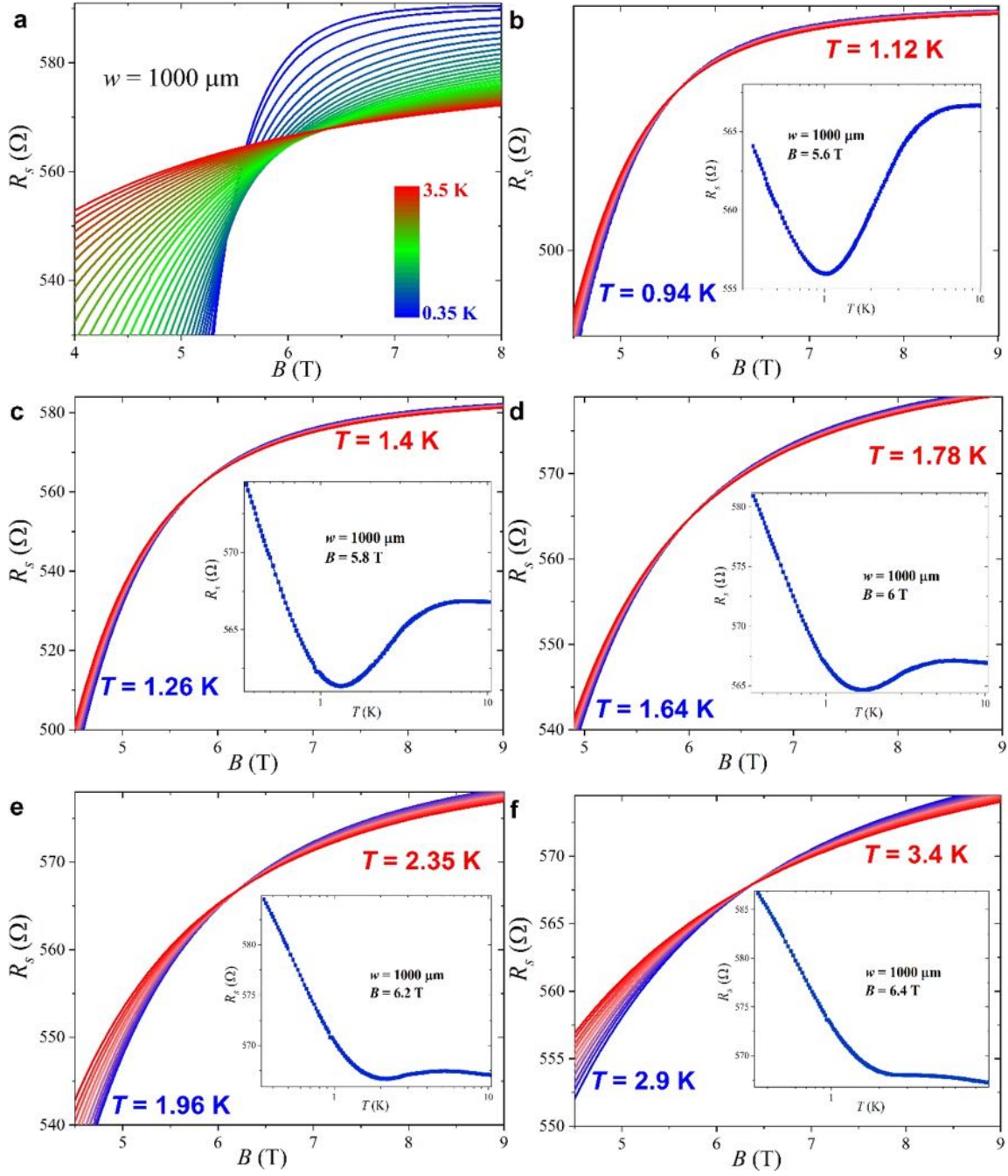

**Fig. 4 | Superconductor-to-insulator phase boundary for the 1000-µm-wide bridge**. (**a**) The $R_s(B)$ at different temperatures ranging from 0.35 K to 3.5 K. (**b** to **f**) The phase boundary corresponds to the crossing points of the $R_s(B)$ data. Inset: the corresponding critical regime at the $T_{min}(B)$ on the $R_s(T)$ curve. The crossing point at the Cooper-pair localization quantum-critical field at $B_c^1$ is shown in Fig. 4a in the main text. The data around $B = 6.4$ T in (**f**) correspond to the pair-breaking critical field $B_c^2$ near $T_c(0)$.

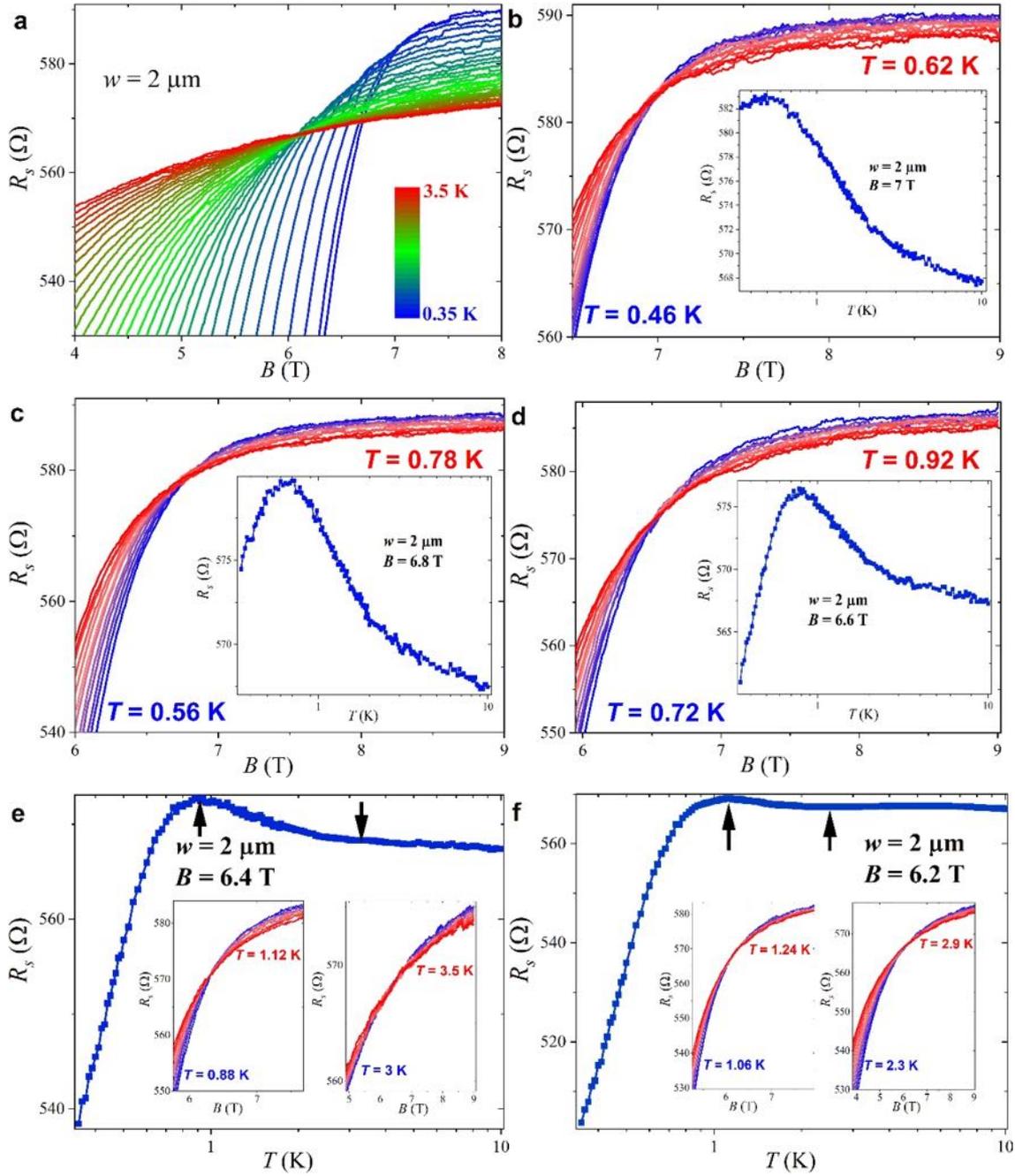

**Fig. 5 | Superconductor-to-insulator phase boundary for the 2-μm-wide bridge**. (**a**) The $R_s(B)$ at different temperatures ranging from 0.35 K to 3.5 K. (**b** to **d**) The phase boundary is defined by the crossing points on the $R_s(B)$ curves above $B_c^2 = 6.4$ T. Inset: the corresponding critical regime at the $T_{max}(B)$ on the $R_s(T)$ curves. (**e** and **f**) The critical regime at the respective $T_{min}(B)$ and $T_{max}(B)$ in magnetic fields below $B_c^2 = 6.4$ T. Inset: the corresponding crossing points in the $R_s(B)$ data.

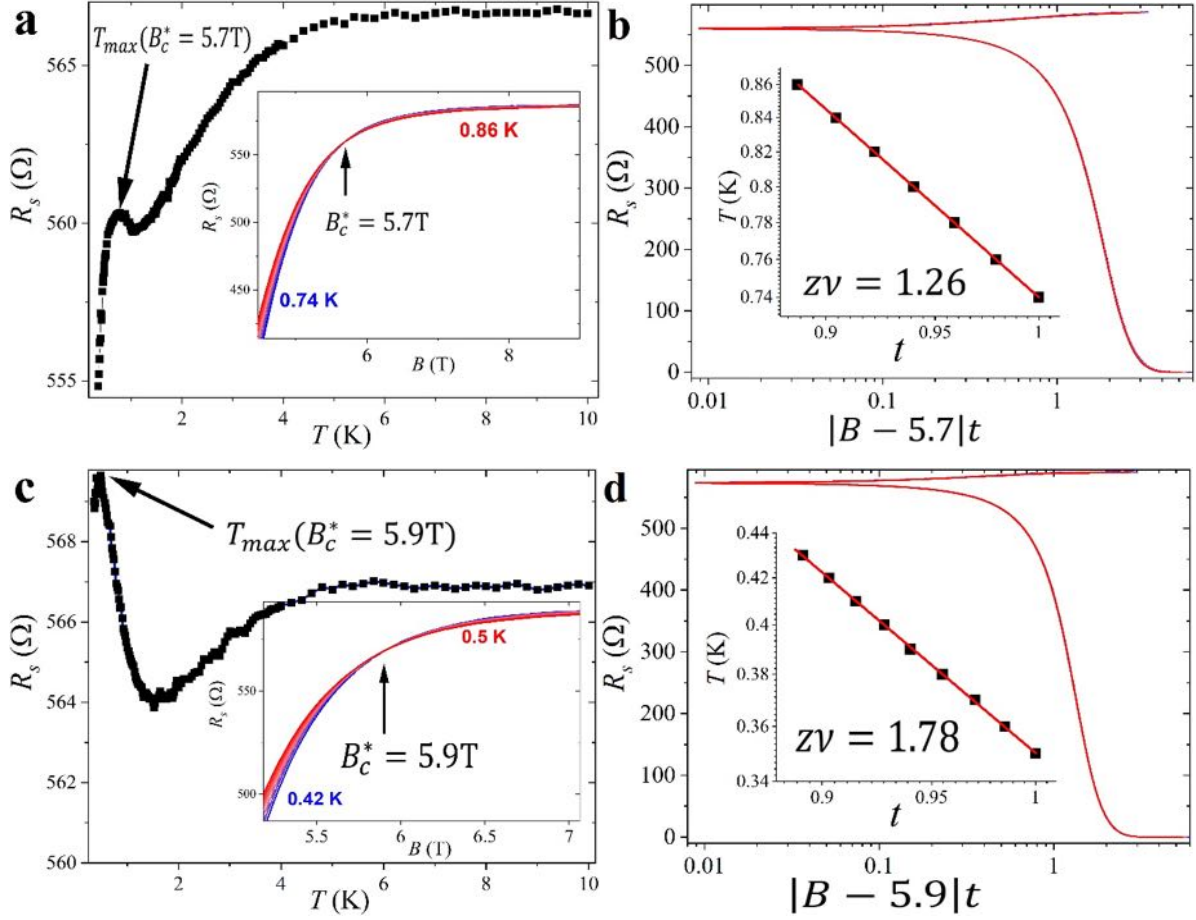

**Fig. 6 | Schematic diagram for the scaling analysis for the superconductor-to-insulator transition at $T_{max}(B_c^*)$.** (**a** and **c**) Sheet resistance as a function of temperature at $B_c^* = 5.7$ T and $5.9$ T, respectively. The corresponding critical regimes are marked by the black arrows. Inset: $R_s(B)$ curves within the corresponding critical regime. (**b** and **d**) Rescaled $R_s(B)$ curves as functions of $|B - B_c^*|t$. Inset: temperature dependence of the scaling parameter $t$.

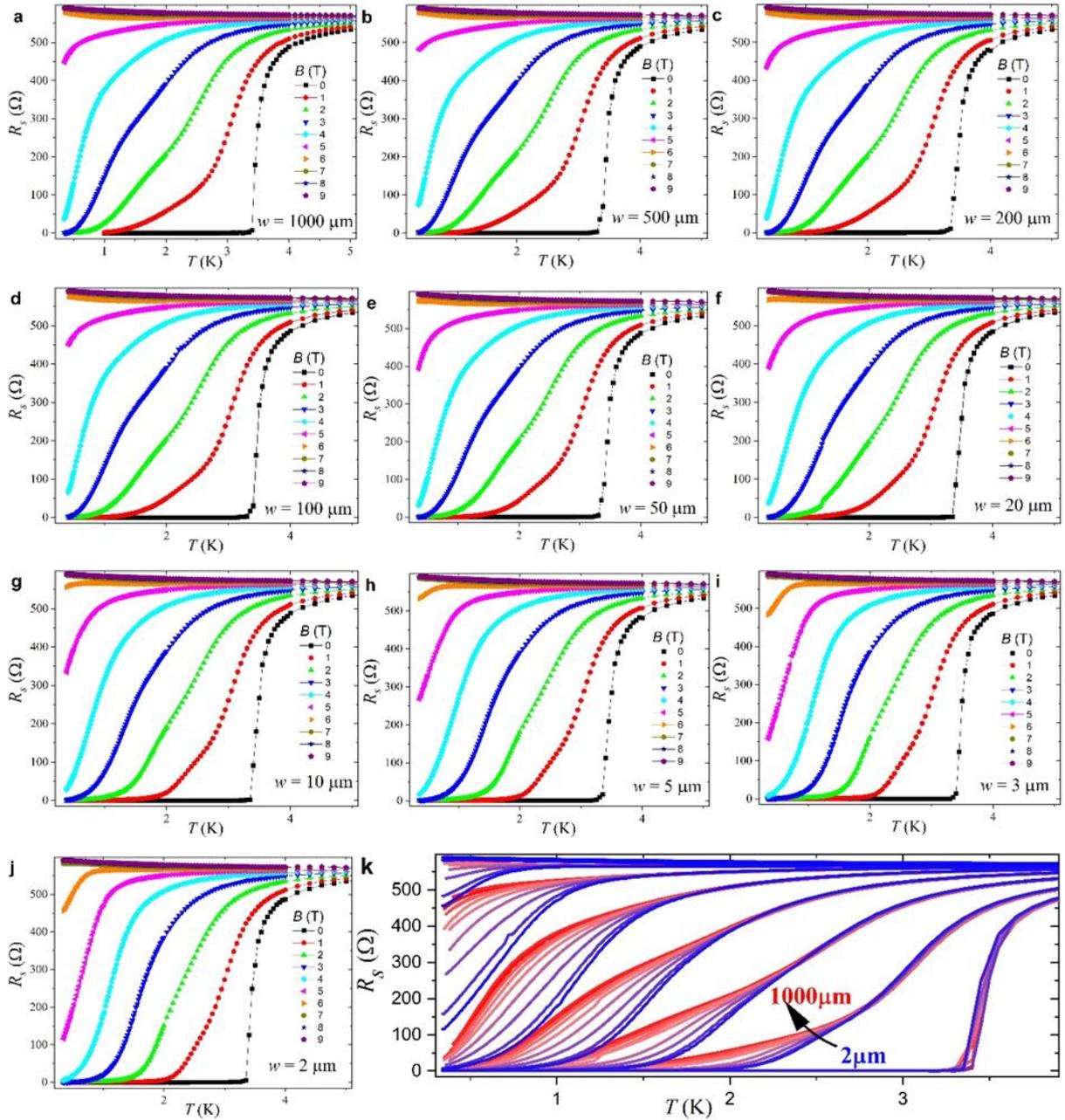

**Fig. 7 | The size effect induced suppression of the resistivity near the normal-to-superconductor transition.** (**a** to **j**) Sheet resistance $R_s$ as functions of temperature in magnetic fields from 0 to 9 T. (**k**) The strong suppression of the resistivity near the normal-to-superconductor transition below the corresponding critical temperature $T_c(B)$ is due to the narrowing of the width of the superconducting bridges well below $\Lambda(0) \approx 350$ μm.

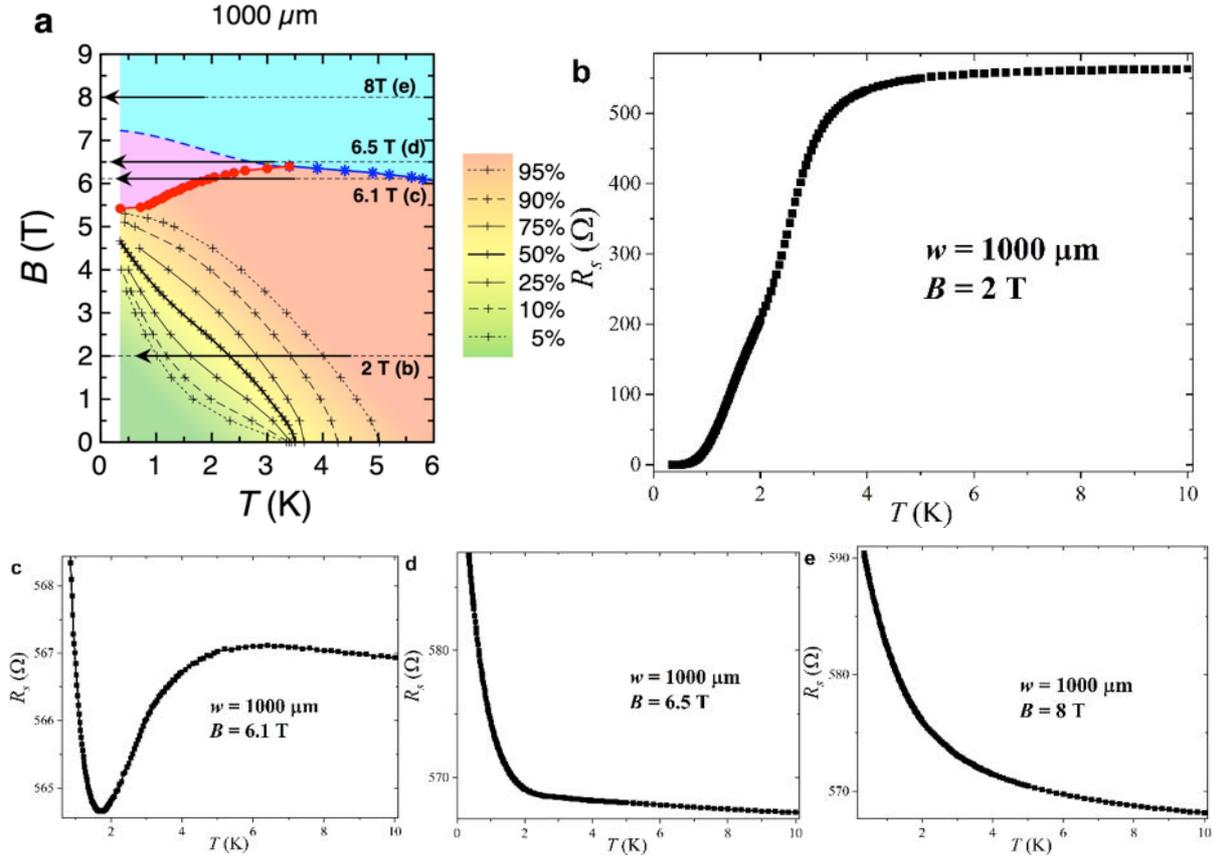

**Fig. 8 | Illustrating diagrams for the magnetic phase diagram of the superconductor-to-Bose insulator transition.** (**a**) Phase diagram of the 1000-μm-wide bridge, representing the phase diagram of infinite 2D superconducting systems, showing sequential superconductor-to-Bose insulator-to-Fermi insulator quantum phase transitions. (**b**) Representative $R_s(T)$ dependence for magnetic fields far below the quantum critical field $B_c^1$. (**c**) Representative $R_s(T)$ dependence for fields between $B_c^1$ and $B_c^2$. (**d**) Representative $R_s(T)$ dependence for fields between $B_c^2$ and $B_c^F$. (**e**) Representative $R_s(T)$ dependence for fields above $B_c^F$. The most significant difference between the Bose-insulating and the Fermi-insulating states (**d** and **e**) is that the $dR_s/dT$ in the Bose-insulating state shows a qualitatively much more pronounced divergence towards zero temperature than that in the Fermi-insulating state.

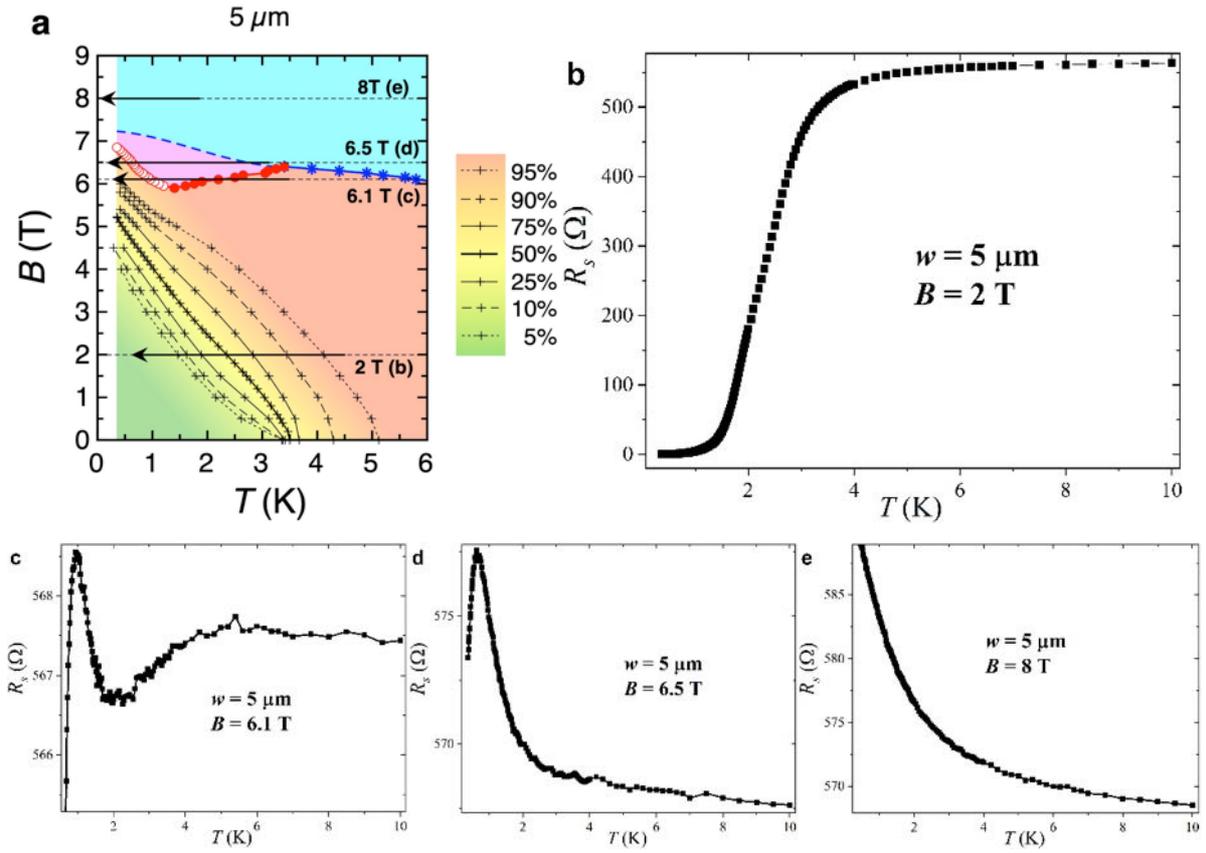

**Fig. 9 | Illustrating diagrams for the magnetic phase diagram in the narrow bridges.** (**a**) Phase diagram of the 5-µm-wide bridge. (**b**) Representative $R_s(T)$ dependence for magnetic fields far below the quantum critical field $B_c^1$. (**c**) Representative $R_s(T)$ dependence for fields between $B_c^1$ and $B_c^2$. (**d**) Representative $R_s(T)$ dependence for fields between $B_c^2$ and $B_c^\star$. (**e**) Representative $R_s(T)$ dependence for fields above $B_c^F$.

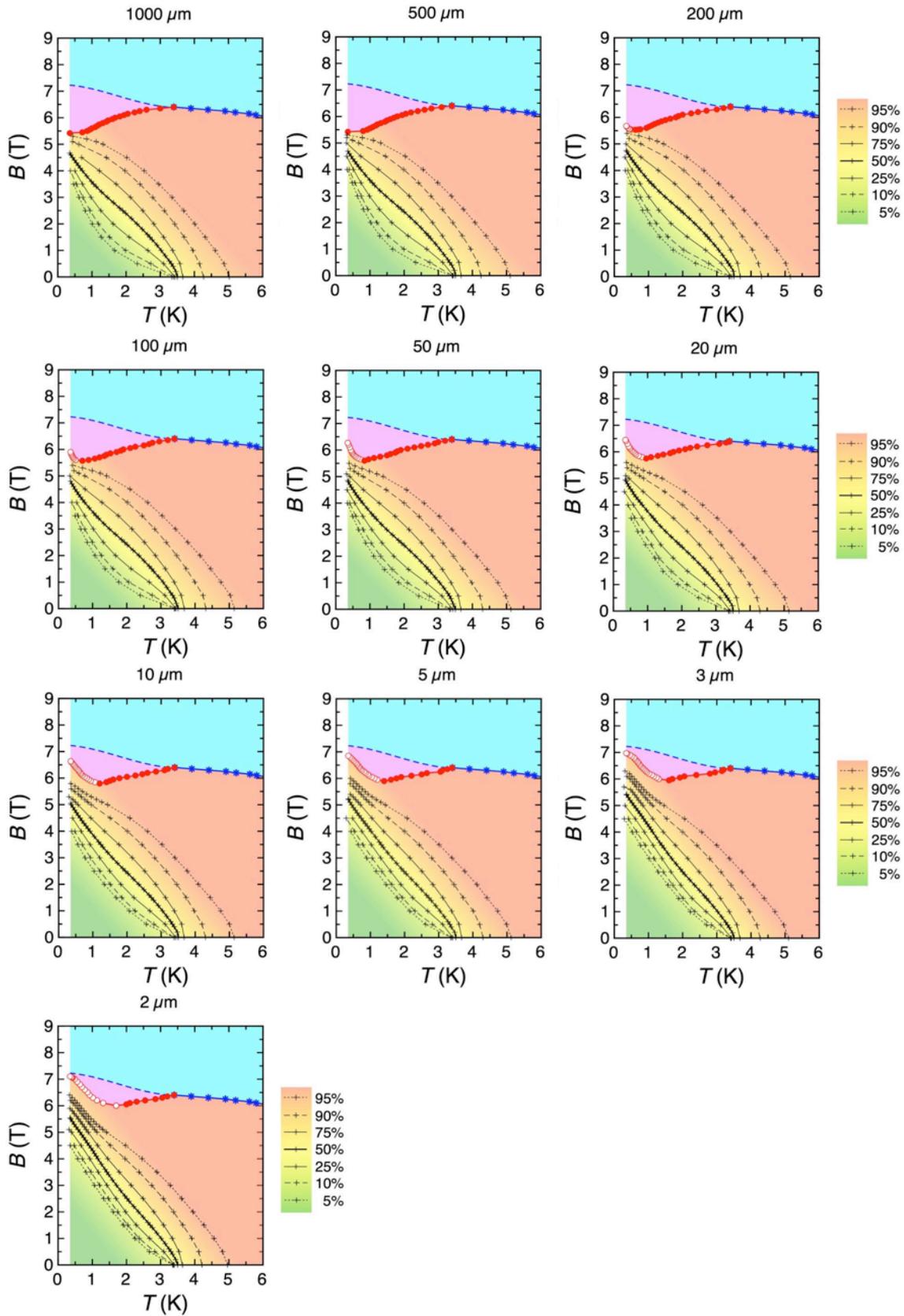

**Fig. 10 | Schematic *B-T* phase diagrams for all the bridges.** For a video showing the evolution of the magnetic phase diagrams as a function of bridge width, see the Online Video 2 on https://www.physik.uzh.ch/groups/schilling/paper/Phasediagrams.mov

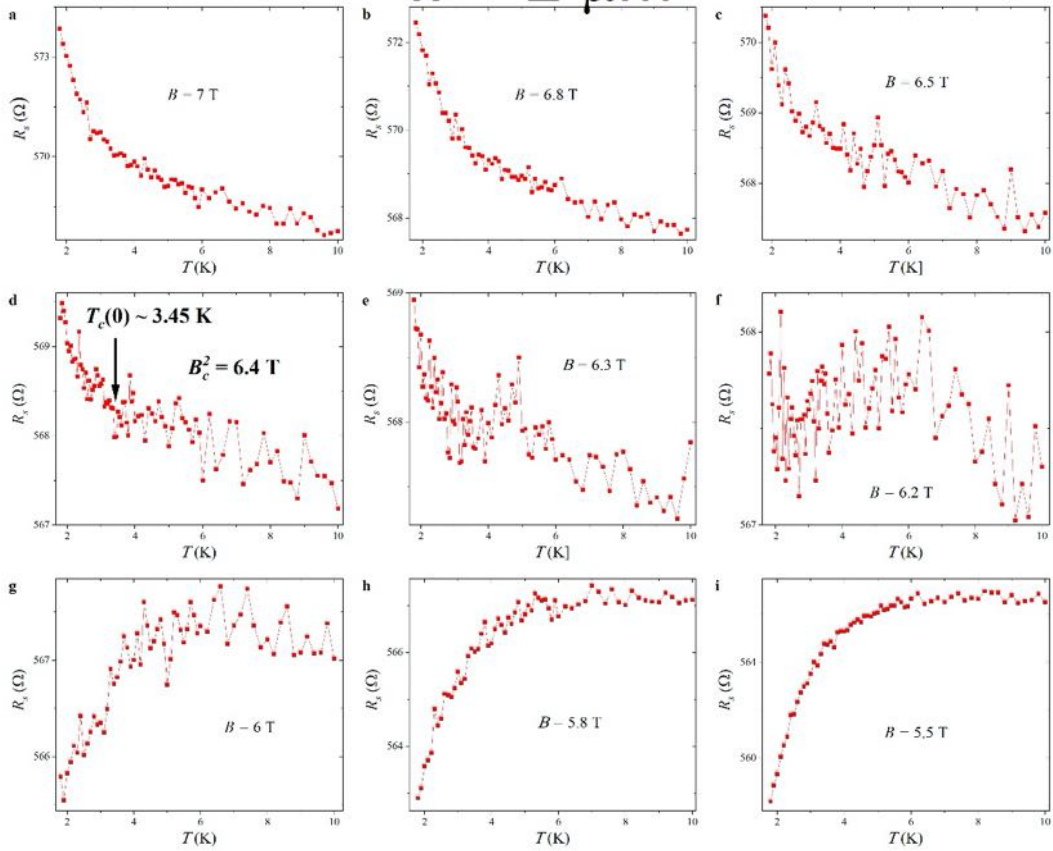

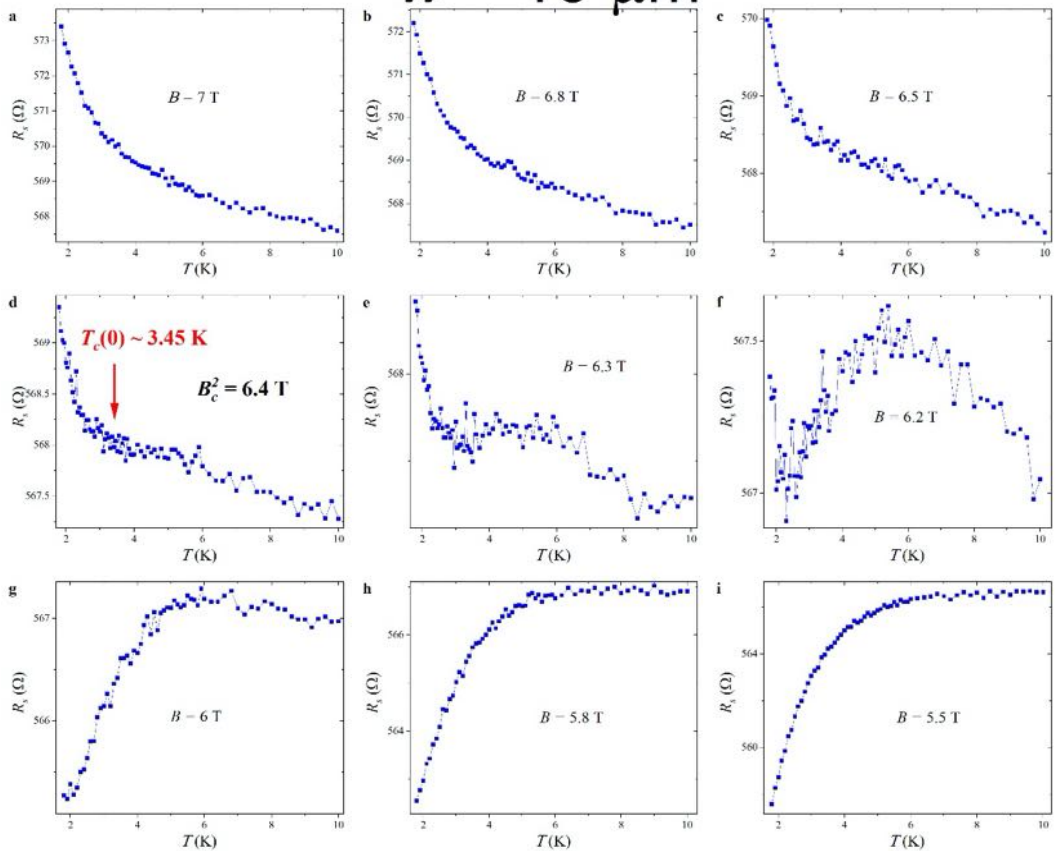

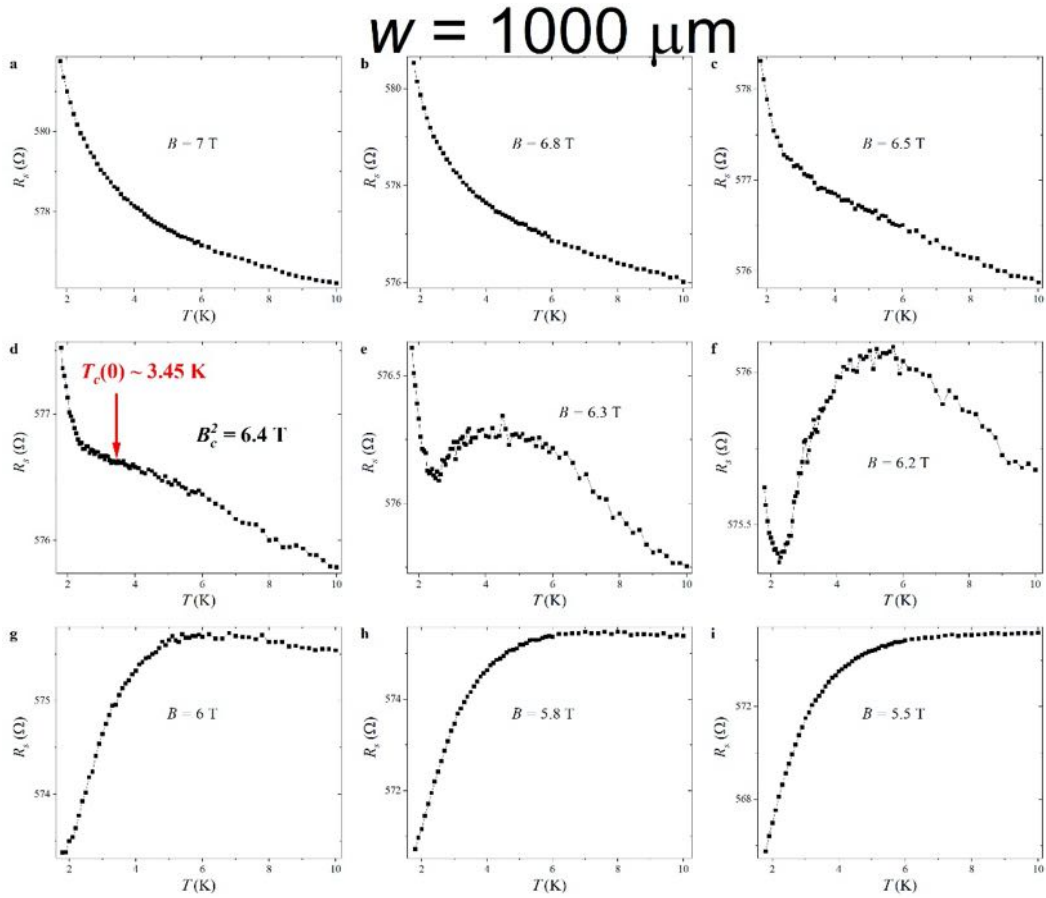

**Fig. 11 | Demonstration of the depairing critical field $B_c^2$ at $T_c(0)$ for the 2-, 10-, and 1000-μm-wide bridges.** (**a** to **c**) Temperature dependence of $R_s$ for magnetic fields far above $B_c^2$. (**d**), Temperature dependence of $R_s$ at $B_c^2$, where a plateau appears on the $R_s(T)$ curve near $T_c(0)$. (**e** to **i**) Temperature dependence of $R_s$ for fields below $B_c^2$, where the Bose-insulating state appears.

## Supplementary Table

**Table 1 | Fitting results from the activated scaling law for all bridges**

| $w$ (μm) | C | $B_c^F$ | $\nu\Psi$ |
|---|---|---|---|
| 2 | $0.792 \pm 0.002$ | $7.25 \pm 0.01$ | $0.573 \pm 0.009$ |
| 3 | $0.829 \pm 0.003$ | $7.25 \pm 0.01$ | $0.632 \pm 0.009$ |
| 5 | $0.925 \pm 0.056$ | $7.25 \pm 0.07$ | $0.919 \pm 0.084$ |
| 10 | $1.463 \pm 0.161$ | $7.26 \pm 0.11$ | $0.981 \pm 0.106$ |
| 20 | $2.228 \pm 0.303$ | $7.26 \pm 0.09$ | $1.533 \pm 0.107$ |
| 50 | $2.494 \pm 0.697$ | $7.24 \pm 0.19$ | $1.582 \pm 0.206$ |

# Online Videos

**Online Video 1** for a clearer visualization of the evolution of $R_s(B,T)$ on the bridge width, see
https://www.physik.uzh.ch/groups/schilling/paper/Resistivity.mov

**Online Video 2** showing the evolution of the magnetic phase diagrams as a function of bridge width, see https://www.physik.uzh.ch/groups/schilling/paper/Phasediagrams.mov